\documentclass[12pt,a4size]{article} 
 
\usepackage[T2A]{fontenc} 
\usepackage{amssymb} 
 
 
\makeatletter 
 
\begin{document} 
 
\begin{center} 
{\LARGE Constraints on the photon charge based on observations of 
extragalactic sources} 
 
\vskip 0.5cm 
 
V.V. Kobychev$^{1,*}$, 
S.B. Popov$^{2,3**}$.\\   
$^1$Institute for Nuclear Research of Ukrainian NAS, \\ 
$^2$University of Padova,  $^3$Sternberg Astronomical Institute. 
 
\end{center} 
 
\vskip 1cm 

{\it Using modern high-resolution observations of extragalactic compact radio
sources  we obtain an estimate of the upper bound on a photon electric
charge at the level $e_{\gamma} \lesssim 3 \cdot 10^{-33}$ of elementary
charge (assuming the photon charge to be energy independent). 
This is three orders of magnitude better than the limit obtained
with radio pulsar timing. Also we set a limit on a photon charge in
the gamma-ray band (energies about 0.1~MeV).
In future the estimate made for extragalactic sources can be significantly
improved.  } 
 
\vskip 0.5cm 
 

\vfill 


$^*$ kobychev@kinr.kiev.ua, $^{**}$ polar@sai.msu.ru 
 
\newpage 
 
\section{Introduction} 

The most restrictive up-to-date upper limit on an electric charge of photons
was obtained from the timing of millisecond radio pulsars. Radio impulses 
are smeared due to a 
dispersion of charged photons moving through the interstellar 
magnetic field (Cocconi, 1988; the result was later refined by Raffelt, 1994):

$$
e_{\gamma }/e<5\cdot 10^{-30}.
$$ 
 
A little bit weaker constraint on a photon charge was discussed by
Cocconi (1992) with a different approach based on an angular spread of photons
propagating from distant extragalactic sources. This spread arises
due to a deviation of a photon having a hypothetic small charge in 
a magnetic field. An estimate of 
 $ e_{\gamma }/e < 10^{-27.7}\approx 2\cdot 10^{-28}$   
has been obtained with this method from an examination of 
a photon trajectory in
the magnetic field of our Galaxy ($B\sim 10^{-6}$ G, 
path length $l\sim 10$ kpc). This restriction can be significantly 
improved by an increasing of the path length 
(i.e. it is necessary to study the effect in extragalactic 
fields) and, additionally, by extending the bandwidth (the limit of 
Cocconi (1992) is based on observations in a quite narrow 
bandwidth  $\sim 2$ MHz).
 
Another constraint has been obtained recently by a study of properties of 
cosmic microwave background. An existence of a small photon charge would 
result in charge asymmetry of the Universe and would contribute to the
observed CMB anisotropy. The quantitative consideration (Caprini {\it et al.},
2003) leads to a very strong upper limit of $e_{\gamma }/e < 10^{-38}$,
but it is valid only in the case of non-anticorrelated charge
asymmetries produced by different types of particles, and, more important,
if  photons have charges of only one sign. These assumptions make 
this limit  model-dependent\footnote{See also an earlier paper by
Sivaram (1994) where the author discusses a limit based on the cosmic
microwave background radiation.}.

We note, that the best laboratory limit  
$e_{\gamma }/e < 8.5\cdot 10^{-17}$ (Semertzidis {\it et~al.}, 2003) is 
significantly worse than the astrophysical restrictions.

\newpage 
 
\section{Calculations and estimates} 

Recent observations show that magnetic fields in many 
galactic clusters are as large as few microgauss with a characteristic
autocorrelation length of few kpc (see a review of Carilli and Taylor 
(2002) and references therein). If a photon exhibits small but finite 
electric charge (and all photons have the same sign of the charge)
then photons with different energies moving
through intracluster magnetic field would follow different
trajectories. It would result in an increase of an angular size of 
a source (of course the same but smaller effect should exist for
intercluster fields). Also in observations at two different frequencies
centers of images would be shifted relative to each other
\footnote{We assume the angles of deviation to be small throughout 
this article.}, simultaneously emitted photons of different energies
would reach an observer at different moments. If photons have
different signs of the charge (but the same absolute values) then images
would be smeared even for monoenergetic radiation; however, 
the position of the image center should not depend on photons energy in
this case.

Let an ultrarelativistic particle (a photon) with an electric charge  
$e_{\gamma }$ and a momentum  $p=h\nu/c$ move through a magnetic field with 
a component $B_{Y}$ orthogonal to the momentum. Its trajectory will
have a curvature radius equal to $cp/(e_{\gamma }B_Y)$. One can see that a
deviation will be more significant for low-energy radio photons, and, 
besides, the angular resolution of radio observations (with the VLBI technique)
is much better than in other parts of electromagnetic spectrum 
-- it can be as small as $10^{-5}$ arcseconds. Thus, one should expect
that the best restriction on a photon charge will be obtained in
the radio frequency range. But taking into account that the origin of the
effective charge of photon can be related to violation of 
Lorentz-invariance we will discuss the upper limit on a photon charge
in a wide range of energies.
 
A photon travelling along an arc with a radius of $r_H$ after passing a 
distance $dl$ turns by an angle  $dl/r_{H}$ (in radians). So  
the source at a distance $l_{\star }$ from a detector will 
be observed shifted by an angle 

\begin{equation} 
\varphi _{X}=\int _{0}^{l_{\star }}\frac{dl}{r_{H}(l)}=\frac{e_{\gamma  
}}{h}\int _{0}^{l_{\star }}\frac{B_{Y}(l)dl}{\nu (l)}. 
\end{equation} 
along the $X$-axis orthogonal to the 
line of sight ($Z$ axis).
 
A dependence of a frequency on $l$ appears for  cosmological 
distances due to a redshift: $\nu (z)=(1+z)\nu _{0}$.
 
Two photons with different energies diverge by an angle
(if the photon charge is not energy dependent): 
 
\begin{equation} 
\Delta \varphi =\varphi _{1}-\varphi _{2} 
= 
\frac{e_{\gamma }}{h} 
\int_{0}^{l_{\star }}B_{Y} 
\left(\frac{1}{\nu _{1}}-\frac{1}{\nu _{2}}\right)dl. 
\end{equation} 
 
Thus, observation of a source with an angular diameter $\Delta \varphi$ in 
a band $\Delta \nu$ (with $\Delta \nu \ll \nu$) leads to
a constraint on a photon charge:

\begin{equation} 
e_{\gamma }/e \lesssim \frac{\Delta \varphi h}{e}\left(\int _{0}^{l_{\star  
}}\frac{B_{Y}\Delta \nu dl}{\nu ^{2}}\right)^{-1}. 
\end{equation} 

On the other hand, when observations are performed in two well separated 
frequency ranges ($\nu_{1} \ll \nu_{2}$) one can use an approximation
$\Delta \nu /\nu^2\approx 1/\nu_1$, and

\begin{equation} 
e_{\gamma }/e \lesssim \frac{\Delta \varphi h}{e}\left(\int _{0}^{l_{\star  
}}\frac{B_{Y} dl}{\nu_1 }\right)^{-1}. 
\end{equation} 
Here $\Delta \varphi$ should be considered as an angular distance between
two images in two different bands. 
The integrals in the equations (3) and (4) can be estimated from an
observational data on the Faraday rotation of polarization plane
of the radio wave propagating in magnetized plasma. This quantity is expressed
in  terms of rotation measure ($RM$) defined as the angle of polarization
rotation divided by the wave length squared. For a completely ionized 
medium we have (Clark {\it et al.}, 2001):

\begin{equation}
RM = 8.12\times10^5 \int  _{0}^{l_{\star}} n_e B_{Z}dl.
\end{equation}

The distance is given in Mpc, $n_e$ in cm$^{-3}$,
$B_{Z}$ is the longitudinal projection of magnetic 
field (in microgausses). In the case of isotropic distribution of
the field, the longitudinal projection under the integral can be
changed by a projection on any other axis, for instance by $B_{Y}$.
Assuming this further we will omit the projection index of $B$.
 
The redshift dependence is neglected in the equations above. In the case 
of $z$ being not negligible, the formulae should be rewritten in the terms 
of the redshift rather than the distance. Beside this, the cosmological 
effects should be taken into account. 
 
The element of integration may be expressed as:
 
$$ 
dl=-\frac{c}{H_{0}}(1+z)^{-3/2}dz 
$$ 
(this equation is related  to the flat Universe without the ``dark energy''
contribution; taking this contribution into account would make
the limit more restrictive). $B(z)=B_{0}(1+z)^{2}$,  $\nu(z)=\nu _{0}(1+z)$ 
(Ryu {\it et al.}, 1998).

Then the Eq.~(3) is transformed to:
 
$$ 
e_{\gamma}/e < \frac{\Delta \varphi h}{e} 
 \left(\int _{0}^{z_{\star }} 
\frac{\Delta \nu_{0}(1+z)} {\nu_{0}^{2}(1+z)^{2}} 
B_{0} (1+z)^{2} 
\frac{c} {H_{0}} (1+z)^{-3/2} dz \right)^{-1} = 
$$ 
\begin{equation} 
= \frac{\Delta \varphi h}{e} 
\frac{H_{0}\nu _{0}^{2}}{cB_{0}\Delta \nu _{0}} \cdot 
\frac{1}{2(\sqrt{1+z_{\star}}-1)}. 
\end{equation} 
For $z_{\star} \ll 1$ the last fraction in 
the Eq.~(7) tends to $1/z_{\star}$.

In the case of extragalactic sources 
usually it is difficult (or impossible) to obtain a good estimate of the 
magnetic field on the line of sight. 
Here for illustrative purposes we derive a limit on a photon electric
charge using modern estimates of large scale extragalactic magnetic fields.
Considering the effect of large scale magnetic field (with the scale 
larger than the size of a galactic cluster) one can use the estimation
made by Kronberg (1994) for the upper limit on the ``cosmologically
aligned'' magnetic field: $B_0 < 10^{-11}$~G (these data are obtained
from the upper bound of 5~rad/m$^2$ on any systematical growth of
$RM$ with distance for $z=2.5$), as well as the upper limit of
$B_0 < 10^{-9}$~G for changing field with the correlation length of
$\sim 1$~Mpc. Widrow (2002) gives the upper limit on the uniform component
of the cosmological field of 
$B_0 < 6\cdot 10^{-12}\textrm{~G~}(n_e/10^{-5}\, \textrm{cm}^{-3})^{-1}$.
This limit agrees with the quoted above estimation by Kronberg.
Different investigations (see the up-to-date review in Widrow, 2002)
indicate that the real rotation measure cannot be less than the given
upper limit by 2-3 orders of magnitude. So we can conservatively 
take $B_0 > 6\cdot 10^{-15}\textrm{~G}$ as the lower limit for 
the ``non-compensated'' cosmological field.

Let us use real observations to estimate an effect of such low magnetic
fields. As part of the VSOP (VLBI Space Observatory program)
Lobanov {\it et al.} (2001) observed the quasar 
PKS~2215+020 at $\nu_0 = 1.6$~GHz in a bandwidth  $\Delta \nu_0 = 32$~MHz.
The angular resolution was about 1 mas. The redshift of the source is
$z_{\star} = 3.57$ that corresponds to the distance  
$l_{\star} \approx 4700$~Mpc.
Substituting these quantities into Eq.~(6) and assuming 
$B_{0}=6\cdot 10^{-15}$~G, $H_{0}=70$~km/s/Mpc=$2.3\cdot 10^{-18}$~1/s,
we get the limit on a photon charge:
$$ 
e_{\gamma }/e \lesssim 6 \cdot 10^{-29}. 
$$ 
It is only an order of magnitude worse than the restriction of Raffelt (1994).
However, we use here only a very conservative upper limit on the uniform 
component of extragalactic magnetic field. As another example let
us consider an improvement of this limit (with the same frequencies and 
angular resolution) if the source is observed through a typical
cluster (relatively close to us, $z \ll 1$). Neglecting the cosmological
effects the integral in (3) is transformed to $(\Delta \nu/\nu^2)\int B dl$.
With an estimate of its value as $Bl = 1\mu\textrm{~G}\cdot\textrm{Mpc}$ 
(the product of typical values of intracluster field and the size of
the central part of cluster) we obtain:
 
$$ 
e_{\gamma }/e < \frac{\Delta \varphi h}{e}\frac{\nu^2}{Bl\Delta \nu} 
=2\cdot 10^{-33}. 
$$ 

Thus, observation of a source through relatively high intracluster
fields (which are, besides, known with better accuracy than the fields 
outside clusters) allows to improve significantly the limit on an electric 
charge of photon in spite of the shorter pass length in the field.

\newpage 
 
\section{Specific example} 

In the example presented below a cluster of galaxies (which works as a 
``scattering screen'') has $z\ll 1$ 
and the influence of an intercluster field is neglected.
When $z$ is low, the dependence of frequency on distance in Eqs. (3)
an (4) can be neglected. Owing to this the limit on a photon charge 
can be expessed directly in  terms of the observable rotation measure
by excluding the distribution of magnetic field end electron density
along the photon trajectory:
 
\begin{equation} 
e_{\gamma }/e \lesssim  
3.2 \cdot 10^{-19} \frac{\Delta \varphi h}{e} 
f(\nu)^{-1} 
\frac{812h_{70}^{1/2}}{RM}, 
\end{equation}
where $f(\nu)=\Delta \nu /\nu^2$ or $1/\nu_1$ if the frequencies
are close or distant correspondingly. We use Eq.~(5) and express
an electron density as $n_e=10^{-3} h_{70}^{1/2} \textrm{1/cm}^{3}$ 
(Clarke {\it et al.}, 2001)\footnote{Note that the normalized
Hubble constant $h_{70}$ 
is always used with a lower index, whereas the Planck constant $h$ is written
without an index.}.

When observations are performed in separated frequencies 
($\nu_1 \ll \nu_2$) this formula can be rewritten as:
 
\begin{equation} 
e_{\gamma}/e = 1.8\cdot10^{-32}  
               h_{70}^{1/2} 
               \left( \frac {\Delta\varphi} {0.001''} \right) 
               \left( \frac {\nu_1} {1\textrm{~GHz}} \right) 
               \left( \frac {RM} {1\textrm{~rad/m}^2} \right)^{-1}. 
\end{equation}

Let us consider the compact source 3C84 in the galaxy NGC1275 that is situated
close to the center of the cluster Abell~426 (the Perseus cluster, $z=0.0183$).
This source was observed by Scott {\it et al.} (2004). These authors made a
survey of 102 active galactic nuclei at 5~GHz with the VSOP facility
(the VLBI network with the space antenna HALCA).  Among six components of
3C84 the smallest one has a diameter (at FWHM) 0.8 mas. Taking into account 
10\% precision cited in that paper we conservatively 
assume the angular diameter to be 0.9 mas. The central frequency and the band
width are 4.8 GHz and 32 MHz correspondingly.

The rotation measure for 3C84 was measured by Rusk (1988):

$$
RM = +76\, \textrm{rad/m}^2.
$$ 

In addition it is worth noting that the Perseus
cluster is a source of polarized dispersed radio emission at 350 MHz 
(Brentjens, de Bruyn, 2003) with $RM\sim 25$-$90$~rad/m$^2$ (including
the cluster outskirts). Taking all together we can safely assume that at
least a rotation measure $\sim 25$ rad/m$^2$ is acquired not in the central
regions with high electron density but in the outer regions of the cluster. 
According to Churazov {\it et al.} (2003) 
$n_e$ outside the central sphere with a radius $\sim 0.3$~Mpc is small
($ \lesssim 10^{-3}\textrm{~cm}^{-3}$)  and 
weakly depends on the distance from the cluster center. 

At first let us discuss the case of two signs of a photon charge.
In this case the smearing of a point source appears even in observations in
a narrow band, so we can use Eq.~(8). Substituting $\nu_1=4.8\textrm{~GHz}$,
$RM=25$~rad/m$^2$,
$n_e=10^{-3}$~1/cm$^3$, $\Delta\varphi=0.9$~mas, we obtain a limit on 
the absolute value of a photon charge:

$$
e_{\gamma }/e \lesssim 3\cdot10^{-33}.
$$

For photons with the one sign of charge, the widening appears due to
different energies of the particles, i.e. it depends on the bandwidth
($\Delta\nu=32\textrm{~MHz}$ in our case), and the effect is smaller.
Using Eq.~(7) and $f(\nu)=\Delta \nu / \nu^2$ we have:

$$ 
e_{\gamma }/e \lesssim 4\cdot10^{-31}. 
$$ 
 
\newpage 
 
\section{Discussion} 
 
Different methods to derive limits on a photon charge are possible.
They may be related to different technique as well as to observations
in different spectral ranges.

A strong constraint can be obtained from the VLBI observations of
close pairs of sources in several frequencies. In this case 
the angular distance between the sources can be measured as accurately
as tens of $\mu$as (Bartel, 2003). Observations of two sources with
different redshifts in several frequencies (like the ones carried out by
Rioja, Porcas, 2003) can give important upper limits.
 
Cocconi (1992) had also obtained some constraints (quite weak) from data 
on an angular dispersion in optical and X-ray ranges:
$e_{\gamma }/e<10^{-25.4}$. He considered the dispersion in magnetic
fields of the Galaxy. The usage of modern data on extragalactic fields
can provide a significantly improved limit.

Study of gamma-ray bursts with known redshifts (these data were not
available at the time of publication of Cocconi, 1988, 1992 and 
Raffelt, 1994) cannot provide limits comparable with those obtained
from the pulsar timing and from angular deflection of radio sources.
However, the possible energy dependency of a photon charge 
(as mentioned above) gives a good occasion to discuss  
charge limits in a wide energy range.
 
Taking into account that the dispersion for gamma quanta in the
interstellar medium is negligible, the time delay is written as
(Barbiellini, Cocconi, 1987):

$$ 
\Delta t= \frac{e_{\gamma }^2 B^2 l_{\star}^3 }{24 c E^2}. 
$$ 
Here the delay is calculated relative to the arrival time of the photons
with energies much larger than $E$. If it is not the case
(for example observations are made in a narrow band  $\Delta E \ll E$) then 
the delay can be written as: 

$$
\Delta t= \frac{e_{\gamma }^2 B^2 l_{\star}^3 }{12 c E^2}\frac{\Delta
E}{E}.
$$
Both the formulae can be applied to photons with different signs of charge
as well as to photons with the same sign. 

A width of a rising edge of GRB is sometimes shorter than 1~ms
($\sim $200-250~$\mu$s, Schaefer, Walker 1999). This quantity can be taken as 
an estimate of a maximum time delay.
Then for $\Delta E/E=0.5$  we have (neglecting cosmological effects):
 
$$ 
e_{\gamma }/e < 5.6\cdot 10^{-21} \left( \frac{E}{100\, \textrm{keV}} \right) 
\left( \frac{B}{6\cdot 10^{-15} \, \textrm{G}} \right)^{-1} 
\left( \frac{\Delta t}{0.1\, \textrm{ms}} \right)^{1/2} 
\left( \frac{l_{\star}}{1000\, \textrm{Mpc}} \right)^{-3/2}. 
$$
Here we normalize the magnetic field by the lower limit on the uniform
component of the extracluster field without taking into account the 
chaotic component of the field that is not known reliably yet.
As before, the restriction can be strenghtened significantly if a GRB
would be observed through a cluster with known magnetic field.

 
\section{Conclusions} 

Modern VLBI observations of extragalactic radio sources give the
stringest limits on the photon electric charge at the level of
$e_\gamma/e \lesssim 3\cdot 10^{-33}$ (with an assumption that photons with
different signs of the charge are equaly abundant, and that a photon charge
does not depend on energy). These limits can be improved by the VLBI
observations of  close pairs of compact sources through clusters with known
magnetic field as for the case of close sources the precision of angular
distance measurements can be about 10 $\mu$as. Also it is desirable
to use data on several sources to improve the statistics\footnote{This
comment was suggested by an anonymous referee}. 
In future, space radio telescopes
will achieve much better angular resolution (Bartel, 2003; Fomalont, Reid,
2004), so precise observations of extragalactic radio sources will provide
the most restrictive upper limits on the photon electric charge.

\vskip 0.5cm  

The authors thank Andrei Lobanov and an anonymous referee for helpful
comments.  
Special thanks to the organizers of the HEA-2003 conference
and to participants  of the web-project Scientific.Ru.

\end{document}